\documentclass[12pt]{article}

\newcommand{\mypar}[1]{{\bf #1.}}
\newtheorem{theorem}{Theorem}[section]

\usepackage[colorlinks=true,linkcolor=blue,citecolor=blue]{hyperref}

\newcommand{\1}{{\bf \iota}}

\newcommand{\A}{\sigma}

\newcommand{\B}{\pi}

\newcommand{\C}{\eta}

\usepackage{graphicx}
\usepackage{amsmath}

\title{A Weighted Generalization of the Graham-Diaconis Inequality for
Ranked List Similarity}

\author{
  Ali Dasdan \\
  KD Consulting \\
  Saratoga, CA, USA \\
  alidasdan@gmail.com
}

\begin{document}

\maketitle

\begin{abstract}
The Graham-Diaconis inequality shows the equivalence between two
well-known methods of measuring the similarity of two given ranked
lists of items: Spearman's footrule and Kendall's tau. The original
inequality assumes unweighted items in input lists. In this paper, we
first define versions of these methods for weighted items. We then
prove a generalization of the inequality for the weighted
versions. 
\end{abstract}

\section{Introduction}
\label{sec:intro} 

The field of similarity computation is more than a century old, e.g.,
see the review at \cite{DaDa11}; it offers many measures to compute
similarity depending on how the inputs are modelled. In this paper, we
focus on computing the similarity of ranked lists.

Among the many list similarity measures in the literature, Spearman's
footrule and Kendall's tau are commonly used. In this paper, we
generalize these measures to include weights and also to work for
partial lists as well as permutations. The main contribution of this
paper is to prove the equivalence of these weighted versions, by
generalizing a proof by Diaconis and Graham for the unweighted
versions. Here equivalence means these measures are within small
constant multiples of one another.

\section{Related Work}
\label{sec:related}

A detailed related work on similarity in general is given in
\cite{DaDa11}. For more recent work, refer to
\cite{ChLiFe14,FaToMi12,FaMiPu17,KuVa10}. For other ways of
incorporating weights with the similarity measures in question, refer
to \cite{ChLiFe14,FaToMi12,FaMiPu17,KuVa10,Sh98,ShBaTs00}

The work in this paper and that in \cite{DaDa11} were actually part of
a comprehensive multi-year project on all forms of similarity for web
search metrics at Yahoo! Web Search; we started the project around
2007. An internal version of \cite{DaDa11} from the year 2009 with our
proof was cited in \cite{KuVa10}.

\section{Rank Assignment}
\label{sec:rank-def}

Consider the ordered or ranked lists $\A=(a,b,c,d,e,f)$ and
$\B=(b,f,a,e,d,c)$. These lists are {\em full} in that they are {\em
  permutations} of each other, which in turn means these lists contain
the same items or elements, possibly in a different order. The rank of
an element $i$ in $\A$ gives its order and is denoted by $\A(i)$. For
example, the rank of $e$ in $\A$ is $\A(e)=5$ whereas its rank in $\B$
is $\B(e)=4$.

Now consider the ordered lists $\A=(a,b,c)$ and $\B=(b,e,c,f)$. These
lists are {\em partial} in that they are not permutations of each
other. This means one list may contain elements that the other list
does not. Moreover, their sizes may be different too. For example, $a$
is present in $\A$ but absent from $\B$; also, $\A$ has a length of
$3$ whereas $\B$ has a length of $4$. For each list, the rank of an
element in that list is well defined.

For comparing ranked partial lists, it is convenient to consider that
these partial lists are actually permutations of each other, with
missing elements somehow added at the end of each partial list. The
motivation for this view comes from web search engines in that a
search result that is missing from the first page of shown (usually
10) results is most probably still in the search index, which is huge,
but not shown.

To figure out how to add the missing elements, consider again our
example partial lists $\A=(a,b,c)$ and $\B=(b,d,c,e)$. Completing
these lists means having each list contain all the elements from their
union, while preserving the rank of their existing elements. The union
of $\A$ and $\B$ is equal to $\{a,b,c,d,e\}$, where we use the set
notation to imply unordering. Comparing these lists to their union
shows that $\A$ is missing $\{d,e\}$ whereas $\B$ is missing
$\{a\}$. Let $\A'$ denote the completion of $\A$. Here we have two
options for $\A'$: 1) $\A'=(a,b,c,d,e)$ or 2) $\A'=(a,b,c,e,d)$. For
$\B$, we have a single option: $\B'=(b,d,c,e,a)$. In the
literature~\cite{FaKuSi03}, another option is to put $d$ and $e$ at
the same rank; we will not consider it here due to the same search
engine analogy above, i.e., if the search engine returns more results,
they will again be ranked.

Now consider the similarity between $\A'$ and $\B'$. Since $d$ is
before $e$ in $\B$ as well as $\B'$, the first option for $\A'$
increases the similarity whereas the second option decreases the
similarity. Either option can be used for completing partial lists to
permutations; our choice in this paper is the first option yet again
due to the same search engine analogy: Assuming a competitor search
engine has the same elements at the same rank will provide motivation
to speed up innovation to beat the competition.

Now that we know how to transform partial lists to become full lists,
we will focus only on full lists in the sequel. We will do one more
simplification in that without loss of generality, we represent
the lists using the ranks of their elements rather than the elements
themselves. For example, $\A=(a,b,c,d,e,f)$ and $\B=(b,f,a,e,d,c)$
become $\A=(1,2,3,4,5,6)$ and $\B=(2,6,1,5,4,3)$. Without loss of
generality, we could also have used $\B$ as the reference list to
determine the integer names of the elements in $\A$. 

\section{Weighted Measures}

We now define the weighted versions of Spearman's footrule and
Kendall's tau. We also provide examples.

Note that the normalized forms of these measures map to the interval
$[0,1]$ where $0$ means the two input lists are ranked in the same
order and $1$ means the two input lists are ranked in the opposite
order. If a mapping to $[-1,1]$ is desired, where $1$ and $-1$
indicate the same and opposite orders respectively, the normalized
value $v$ needs to be transformed to $1-2v$.

\begin{figure}
  \centering
  \includegraphics[width=0.6\textwidth]{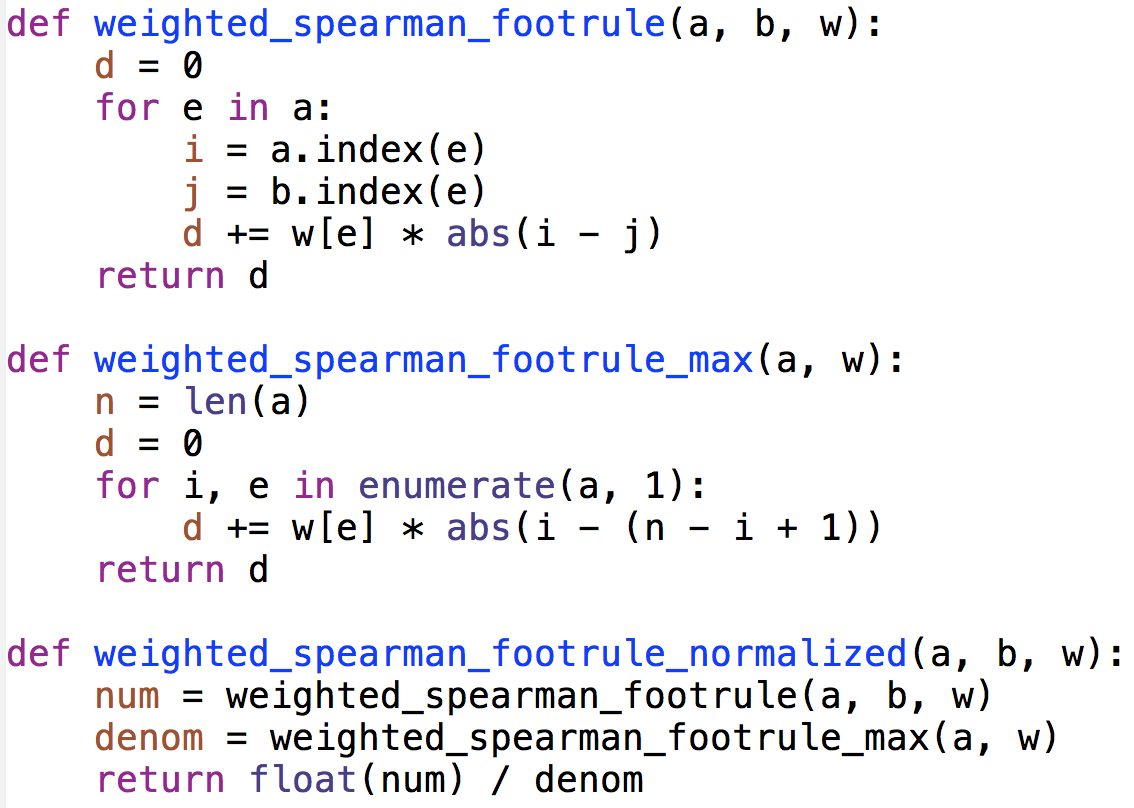}
  \caption{Weighted Sperman's footrule in the Python programming
    language. Use unity weights to derive the unweighted versions.}
  \label{fig:spearman}
\end{figure}

\subsection{Weighted Spearman's Footrule}
\label{sec:footrule}

We define the weighted version of Spearman's footrule
\cite{DiGr77,Spearman1906} for lists of length $n$ as
\begin{equation}
\label{eq:footrule}
S_w(\A,\B) = \sum_{i\in\A\cup\B} w(i) |\A(i) - \B(i)| .
\end{equation}
where $w(i)$ returns a positive number as the weight of the element
$i$.

The measure $S_w$ can be normalized to the interval $[0,1]$ as
\begin{equation}
\label{eq:norm-s-w}
s_w(\A,\B) = \frac{S_w(\A,\B)}{\sum_{i=1}^{n} w(i) |(i) - (n - i + 1)|} 
\end{equation}
where the denominator reaches its maximum when both lists are sorted
but in opposite orders.

Fig.~\ref{fig:spearman} shows the algorithms, written in the Python
programming language, to compute both the numerator and the
denominator of $s_w(\A,\B)$ in Eq.~\ref{eq:norm-s-w} as well as
$s_w(\A,\B)$ itself as their ratio. Using unity weights lead to the
unweighted versions of these algorithms.

\begin{figure}
  \centering
  \includegraphics[width=0.6\textwidth]{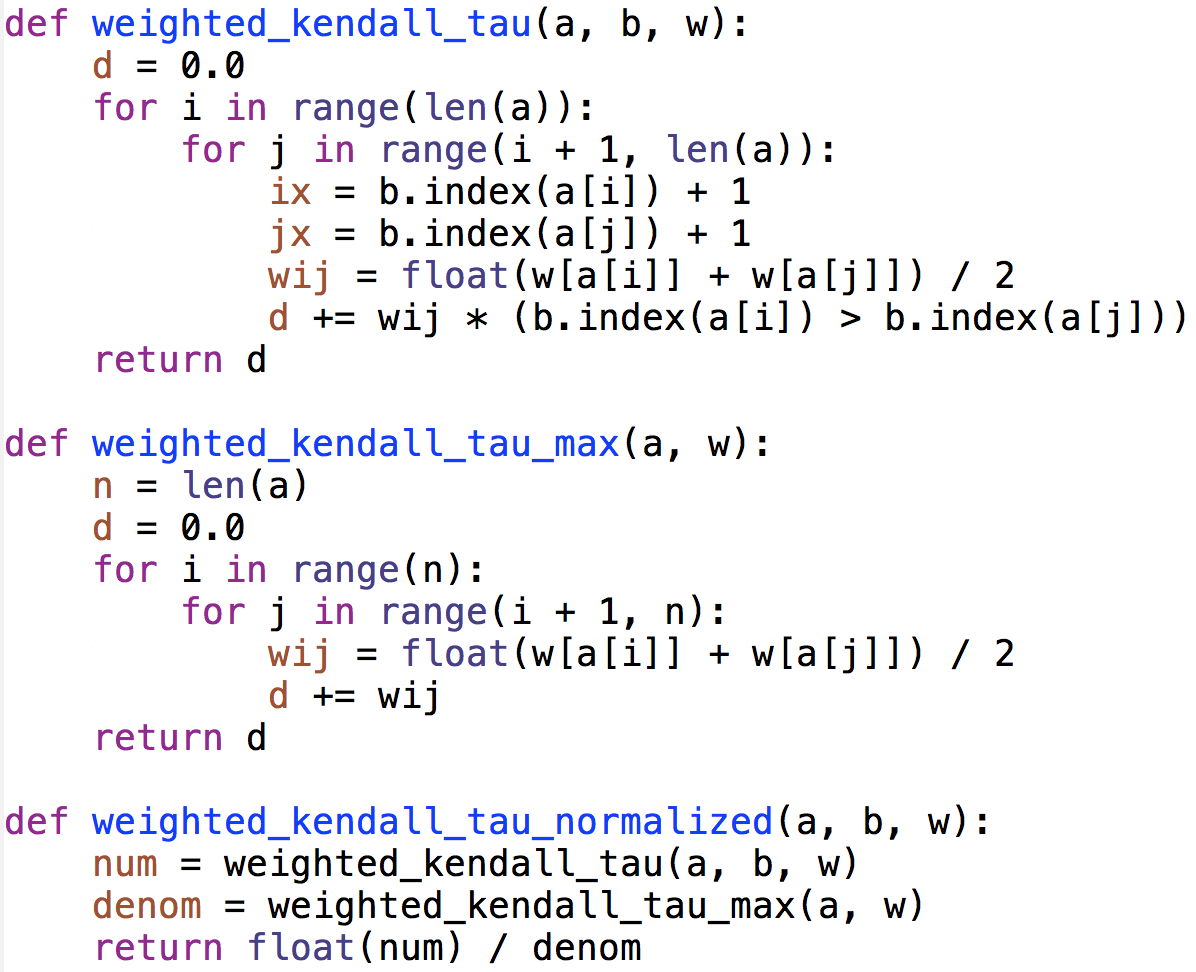}
  \caption{Weighted Kendall's tau in the Python programming
    language. Use unity weights to derive the unweighted versions.}
  \label{fig:kendall}
\end{figure}

\subsection{Weighted Kendall's Tau}
\label{sec:kendall}
The unweighted Kendall's tau is the number of swaps we would perform
during the bubble sort to reduce one permutation to another. As we
described the way we determine the ranks of the extended lists
(\S~\ref{sec:rank-def}), we can always assume that the first list $\A$
is the identity (increasing from 1 to $n$), and what we need to
compute is the number of swaps to sort the permutation $\B$ back to
the identity permutation (increasing). Here, a weight will be
associate to each swap.

We define a weighted version of Kendall's tau
\cite{Kendall1938,Sievers1978} for lists of length $n$ as
\begin{equation}
  \label{eq:kendall}
  K_w(\A=\iota,\B)=\sum_{1\leq i<j\leq n} \frac{w(i)+w(j)}{2}[\B(i)>\B(j)]
\end{equation}
where $[x]$ is equal to 1 if the condition $x$ is true and 0
otherwise.

The measure $K_w$ can be normalized to the interval $[0,1]$ as
\begin{equation}
\label{eq:norm-k-w}
k_{w} = \frac {K_w(\A,\B)}{\sum_{\{i,j\in\A\cup\B:i<j\}} \frac{w(i)+w(j)}{2}} 
\end{equation}
where the value of the denominator is exactly the maximum value that
the numerator can reach when both lists are sorted but in opposite
orders.

Fig.~\ref{fig:kendall} shows the algorithms, written in the Python
programming language, to compute both the numerator and the
denominator of $k_w(\A,\B)$ in Eq.~\ref{eq:norm-k-w} as well as
$k_w(\A,\B)$ itself as their ratio. Using unity weights lead to the
unweighted versions of these algorithms.

\subsection{Examples}
\label{sec:examples}

For the sake of simplicity, let us assume that $w(i)=w$ for all $i$ in
this section. We will have three examples. Let us compute $S_w$,
$K_w$, $s_w$, and $k_w$ for each example.

\mypar{Example 1} Given $\A=(a,b,c,d,e)$ and $\B = (a,b,c,d,e)$, we have
$\A' = (a,b,c,d,e) \sim (1,2,3,4,5)$ and $\B' = (a,b,c,d,e) \sim
(1,2,3,4,5)$. Then,
\[
S_w = 0w=0\mbox{ and }K_w = 0w=0.
\]

Also,
\[
s_w = \frac{0w}{12w} = 0\mbox{ and }k_w = \frac{0w}{10w} = 0.
\]

\mypar{Example 2} Given $\A=(a,b,c,d,e)$ and $\B = (e,d,c,b,a)$, we have
$\A' = (a,b,c,d,e) \sim (1,2,3,4,5)$ and $\B' = (e,d,c,b,a) \sim
(5,4,3,2,1)$. Then,
\begin{align*}
    S_w =& w(|1-5| + |2-4| + |3-3| + |4-2|+ |5-1|)\\
       =& w(4+2+0+2+4)\\
       =& 12w
\end{align*}
and
\begin{align*}
  K_w =& w([5 > 4] + [5 > 3] + [5 > 2] + [5 > 1]\\
      & + [4 > 3] + [4 > 2] + [4 > 1]\\
      & + [3 > 2] + [3 > 1] + [2 > 1])\\
    =& 10w.
\end{align*}

Also,
\[
s_w = \frac{12w}{12w} = 1\mbox{ and }k_w = \frac{10w}{10w} = 1.
\]

\mypar{Example 3} Given $\A=(a,b,c)$ and $\B=(b,d,c,e)$, we first extend
them to full lists and replace elements by their ranks with $\A$ as
the reference list: $\A'=(a,b,c,d,e) \sim \A'=(1,2,3,4,5)$ and
$\B'=(b,d,c,e,a) \sim \B'=(2,4,3,5,1)$. Then,
\begin{align*}
    S_w =& w(|1-2| + |2-4| + |3-3| + |4-5|+ |5-1|)\\
    =& w(1 + 2 + 0 + 1 + 4)\\
    =& 8w
\end{align*}
and
\begin{align*}
    K_w =& w([2 > 1] + [4 > 3] + [4 > 1] + [3 > 1] + [5 > 1])\\
    =& w(1 + 1 + 1 + 1 + 1)\\
    =& 5w,
\end{align*}
where all the other comparisons return zero.

Then the normalized measures become
\[
s_w = \frac{8w}{12w} = \frac{2}{3}\approx 0.67\mbox{ and }k_w = \frac{5w}{10w} = \frac{1}{2}=0.5.
\]

Notice that $K_w\leq S_w \leq 2K_w$ because $5w\leq 8w\leq 10w$.

\section{Equivalence Between Measures}
\label{sec:equivalence}

We now prove that the Graham-Diaconis inequality between Spearman's
footrule and Kendall's tau is valid for weighted ranked lists
too. This inequality shows that these measures with or without weights
are within small constant multiples of each other, which is another
way of saying that these measures are equivalent~\cite{FaKuSi03}.

Denote by $S$ and $K$ the unweighted versions of $S_w$ and $K_w$,
respectively. The unweighted versions are obtained by setting each
weight in $S_w$ and $K_w$ to unity. This equivalence proof allows us
to use the simpler of these two measures, Spearman's footrule, as our
list similarity measure even for the weighted case.

For permutations, the equivalence between the unweighted versions $S$
and $K$ is well known from the following classical result~\cite{DiGr77}:
\begin{theorem}(Diaconis-Graham)
\label{th:DG77}
  For every two permutations $\A$ and $\B$
\begin{equation}
    K(\A,\B)\leq S(\A,\B)\leq 2K(\A,\B) .
\end{equation}
\end{theorem}
For a discussion on the equivalence of other unweighted list
similarity measures, see \cite{FaKuSi03}.

For weighted permutations, we generalize this result to the
equivalence between $S_w$ and $K_w$.

\begin{theorem}
  \label{th:DDperm}
  For every two permutations $\A$ and $\B$, 
  \begin{equation}
    K_w(\A,\B)\leq S_w(\A,\B)\leq 2K_w(\A,\B) 
  \end{equation}
  where every weight is a positive number.
\end{theorem}

Our proof below closely follows the notation and reasoning of the
original proof in \cite{DiGr77} and extends it to the weighted case.

\mypar{Proof} Before proving this theorem, we need the following
  preliminary facts, which are the same as in \cite{DiGr77}. Note that
  the element weights do not invalidate these facts.

  Assume that all permutations below are defined on the same set,
  where a permutation is a bijection (i.e., one-to-one and onto
  function) from some set $X$ of size $n$ to $\{1, \dots, n\}$.

  \mypar{Metric space} Both $S$ and $K$ are metrics, meaning that if
  $d$ denotes one of these measures, then $d$ satisfies the metric
  properties: $d(\A,\B)\geq 0$ (i.e., non-negativity); $d(\A,\B)=0$ if
  and only if (iff) $\A=\B$ (i.e., identity of indiscernible);
  $d(\A,\B)=d(\B,\A)$ (i.e., symmetry); and, $d(\A,\B)\leq
  d(\A,\C)+d(\C,\B)$ for some permutation $\C$ (i.e., triangle
  inequality).

  \mypar{Right invariance} Both $S$ and $K$ are right invariant,
  meaning if $d$ denotes one of these measures, then $d$ is right
  invariant if a permutation $\C$ exists such that
  $d(\A,\B)=d(\A\C,\B\C)$. In particular,
  $d(\1,\A)=d(\A^{-1},\1)=d(\1,\A^{-1})$ where $\1$ stands for the
  identify permutation on $X$ and $\A^{-1}$ is the permutation inverse
  to $\A$ (i.e., $\A\A^{-1}=\1$). To simplify the notation, we
  abbreviate $d(\1,\A)$ by $d(\A)$, hence, $d(\A)=d(\A^{-1})$.

  Now we come to the proof of the theorem. We divide the proof into
  two parts. We prove first $S_w(\A)\leq 2K_w(\A)$ and then $K_w(\A)
  \leq S_w(\A)$.

  \mypar{The proof of $S_w(\A)\leq 2K_w(\A)$} Recall that
$K(\A)=K(\A^{-1})$ is the smallest number of pairwise adjacent
transpositions or swaps required to bring $\A$ to the identity
$\1$. Note that bubble sort will make the same number of swaps to sort
its input list. Let $x_i$, $1\leq i\leq K(\A)$, be a sequence of
integers that indexes a sequence of transpositions transforming $\1$
to $\A_1$ to $\A_2$ to $\dots$ to $\A$. The $i$-th transposition
transforms $\A_i$ to $\A_{i+1}$ by interchanging $\A_i(x_i)$ to
$\A_i(x_i+1)$, i.e., the element at index $x_i$ and the next element
at index $x_i +1$. Without loss of generality, we may assume that
$\A_i(x_i)<\A_i(x_i+1)$. Consider the difference
  \begin{align}
    \label{eq:delta}
    \Delta_{i+1}=&S_w(\A_{i+1})-S_w(\A_i)\\
    =&(w(x_i+1)|x_i-\A_i(x_i+1)|+w(x_i)|x_i+1-\A_i(x_i)|)\nonumber\\
    -&(w(x_i)|x_i-\A_i(x_i)|+w(x_i+1)|x_i+1-\A_i(x_i+1)|)\nonumber
  \end{align}
  where the contributions from the elements whose positions have not
  changed, i.e., the elements at indices other than $x_i$ and $x_i+1$,
  cancel out.
  
  There are three possibilities, each of which removes the absolute
  values in Eq.~\ref{eq:delta} to compute the difference.
  \begin{itemize}
  \item {\em Case 1.} If $\A_i(x_i)<\A_i(x_i+1)\leq x_i < x_i + 1$, then
    $\Delta_{i+1}=-w(x_i+1)+w(x_i)$.
  \item {\em Case 2.} If $x_i<x_i+1\leq \A_i(x_i)<\A_i(x_i+1)$, then 
    $\Delta_{i+1}=w(x_i+1)-w(x_i)$.
  \item {\em Case 3.} If $\A_i(x_i)\leq x_i<x_i+1\leq\A_i(x_i+1)$, then 
    $\Delta_{i+1}=w(x_i+1)+w(x_i)$.
  \end{itemize}
  Thus $S_w(\A)=\sum_{i=1}^{K(\A)}\Delta_i\leq 2K_w(\A)$ because
  $\Delta_{i+1}\leq 2((w(x_i+1)+w(x_i))/2)$ for each of the three
  cases above. Note that the expression of $\Delta_{i+1}$ also
  explains our choice of the additive weight aggregation in $K_w$.
  
   \mypar{The proof of  $K_w(\A) \leq S_w(\A)$}
  To prove the left-hand side of the inequality, first denote the
  inversion $\A(i)>\A(j)$ with $i<j$ by $[i;j]$. We then simplify
  Eq.~\ref{eq:kendall} for $K_w(\A)$ as
  \begin{equation}
    \label{eq:Kw2}
    K_w(\A)=\sum_{1\leq i<j\leq n} \frac{w(i)+w(j)}{2}[\A(i)>\A(j)]\nonumber
  \end{equation}
  or
  \begin{align}
    \label{eq:Kw}
    2K_w(\A) & = & \sum_{1\leq i<j\leq n} w(i)[\A(i)>\A(j)] +\nonumber\\
    &   & \sum_{1\leq i<j\leq n} w(j)[\A(i)>\A(j)],
  \end{align}
  where $[x]$ is equal to 1 if the condition $x$ is true and 0
  otherwise. Note that $2K_w$ is the sum of all inversions $[i;j]$ in
  $\A$ and the weight for each inversion is added twice, once for
  $\A(i)$ and another for $\A(j)$. For the proof, we will show that
  $S_w$ upper-bounds the total number of inversions, hence, each term
  in $2K_w$ separately.
  
  Define two types of inversions: Type I and Type II. Call an inversion
  $[i;j]$ a Type I inversion if $\A(i)\geq j$ and a Type II inversion if
  $\A(i)\leq j$. Note that every inversion of $\A$ is either a Type I or
  a Type II inversion or both. Similarly, the sum of Type I and Type II
  inversions in $\A$ upper-bounds $K_w(\A)$.

  For a fixed $i$, if $[i;k]$ is a Type I inversion, then we must have
  $i<k\leq \A(i)$. Thus, the number of Type I inversions, or the
  number of possible $k$, is at most $\A(i)-i$ and the total weight is
  at most $w(i)(\A(i)-i)$. Similarly, if $[k;j]$ is a Type II
  inversion, then we must have $\A(j)<\A(k)\leq j$. Thus, the number
  of Type II inversions, or the number of possible $k$, is at most $j
  - \A(j)$ and the total weight is at most $w(i)(j-\A(j))$. Then, it
  follows that the first term in Eq.~\ref{eq:Kw} is at most the sum of
  the number of Type I inversions and the number of Type II
  inversions, which is further upper-bounded as
  \begin{align}
    \label{eq:temp}
    \sum_{\A(i)\geq i} w(i)(\A(i)-i)+ \sum_{\A(j)\leq j} w(i)(j-\A(j)) & =
    & \nonumber\\
    \sum_{i} w(i)|\A(i)-i| & = & S_w(\A)\nonumber.
  \end{align}
  
  Similarly, for a fixed $j$, an argument similar to the case for $i$
  and $w(i)$ can be carried out for the case for $j$ and $w(j)$ to prove
  that the second term in Eq.~\ref{eq:Kw} is at most
  $S_w(\A)$. Combined, these two arguments show that $2K_w(\A)\leq
  2S_w(\A)$ or $K_w(\A)\leq S_w(\A)$.
QED.

We generalize this result to the equivalence between $S_w$ and $K_w$
for ranked partial lists.
\begin{theorem}
\label{th:DDpart}
For partial lists $\A$ and $\B$ with the rank assignment for missing
elements as described in \S~\ref{sec:rank-def}.
\begin{equation}
K_w(\A,\B)\leq S_w(\A,\B)\leq 2K_w(\A,\B) 
\end{equation}
where every weight is a positive number.
\end{theorem}

\mypar{Proof} The proof directly follows from the way we assign ranks
of the missing elements in \S~\ref{sec:rank-def}. This is because the
resulting lists become permutations of each other. Thus,
Theorem~\ref{th:DDperm} applies. QED.

\section{Experimental Results}
\label{sec:results}

To provide more insight into these algorithms, we performed two sets
of experiments.

\begin{figure}
  \centering
  \includegraphics[width=0.7\textwidth]{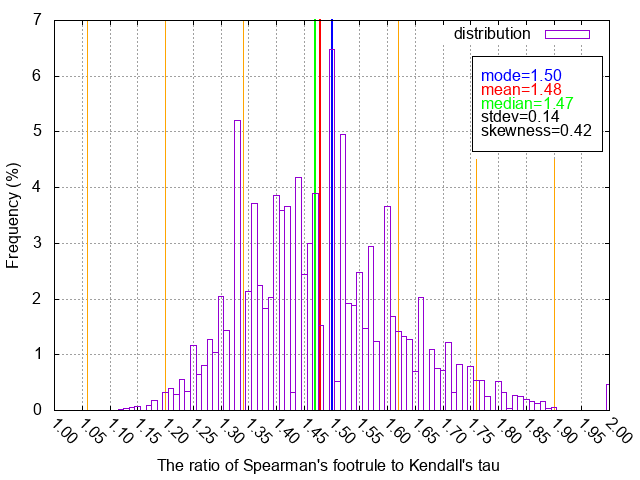}
  \caption{Distribution of the ratio of Spearman's footrule to
    Kendall's tau over all permutations of a 10-element unweighted
    list.}
  \label{fig:dist}
\end{figure}

The first set of experiments is about understanding the distribution
of the ratio of Spearman's footrule to Kendall's tau. We computed this
ratio over all permutations of a 10-element unweighted list. By
Theorem~\ref{th:DG77}, this ratio fits in the range from 1 to 2,
inclusive. Fig.~\ref{fig:dist} shows the outcome distribution with
orange lines marking the multiples of the standard deviation
(0.14). The distribution parameters are also shown in this
figure. Note that the median, mean, and mode are at or very close to
1.50, the middle value of the range; although this may almost imply a
balanced distribution, the distribution is actually slightly
right-skewed with a skewness of 0.42, which is also noticeable
visually.

\begin{figure}
  \centering
  \includegraphics[width=0.7\textwidth]{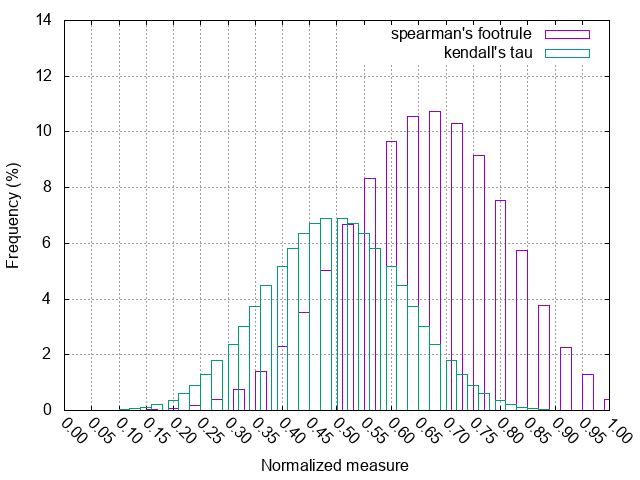}
  \caption{Distribution of the normalized Spearman's footrule and
    Kendall's tau over all permutations of a 10-element unweighted
    list.}
  \label{fig:norm}
\end{figure}

The second set of experiments is about understanding the distributions
of the normalized measures. We again computed the normalized values
over all permutations of a 10-element unweighted list. The normalized
values fit in the range from 0 to 1, inclusive. Fig.~\ref{fig:norm}
shows the outcome distributions for both the normalized measures. As
is also obvious visually, the normalized Kendall's tau is distributed
in a balanced way with the median, mean, and mode are at the middle
value 0.50 of the range (with a standard deviation of 0.13) whereas
the normalized Spearman's footrule is distributed with a negative
skewness (i.e., left-skewed) of -0.18. The other parameters for the
normalized Spearman's footrule are 0.66, 0.66, and 0.68 for the
median, mean, and mode, respectively. The standard deviation is 0.14.

\section{Conclusions}
\label{sec:conclusions}

The field of similarity computation is more than a century old. In
this paper, we focus only on the similarity between two ranked
lists. Such lists may be full (permutations) or partial. For
permutations, the classical Graham-Diaconis inequality show their
equivalence using two well-known similarity measures: Spearman's
footrule and Kendall's tau.

In this paper, we consider ranked lists that may be partial and ranked
lists that may be weighted (with element weights) or both. We first
propose a rank assignment method to convert partial lists to
permutations. Next, we define weighted versions of Spearman's footrule
and Kendall's tau. Finally, we generalize the Graham-Diaconis
inequality to permutations with element weights. Due to the form of
our rank assignment, we also show that the same weighted
generalization applies to partial lists. 
 
\section*{Acknowledgments}

I thank Paolo D'Alberto for a joint work that subsumed this work and
Ravi Kumar for discussions on the weighted form of Kendall's tau.

\bibliographystyle{plain}

\end{document}